\documentclass{esannV2}
\usepackage[dvips]{graphicx}
\usepackage[latin1]{inputenc}
\usepackage{amssymb,amsmath,array}

\begin{document}
\title{Road traffic reservoir computing}

\author{Hiroyasu Ando and Hanten Chang
%
\thanks{This work is partially supported by the grant of JST MIRAI and Kakenhi No. 19K12198, 17H03280, 16H03118, 18K03433, and the collaborative research between University of Tsukuba and Toyota Motor Corporation.}
%
\vspace{.3cm}\\
%
Faculty of Engineering, Information and Systems, University of Tsukuba, \\
1-1-1 Tennoudai, Tsukuba, Ibaraki, 305-8573, Japan
%
}

\setlength{\abovedisplayskip}{2pt} 
\setlength{\belowdisplayskip}{2pt}

\maketitle

\begin{abstract}
Reservoir computing derived from recurrent neural networks is more applicable to real world systems than deep learning because of its low computational cost and potential for physical implementation. Specifically, physical reservoir computing, which replaces the dynamics of reservoir units with physical phenomena, has recently received considerable attention. In this study, we propose a method of exploiting the dynamics of road traffic as a reservoir, and numerically confirm its feasibility by applying several prediction tasks based on a simple mathematical model of the traffic flow. 
\end{abstract}

\section{Introduction}
Information processing in the brain is conducted by means of electrical signals in neural networks (NNs). A model of NNs has been studied for more than half a century \cite{amariNN}. Recently, NNs have exhibited impressive performance in difficult tasks owing to the development of learning algorithms and computers. Specifically, it is well-known that multi-layered NNs are suitable for the task of image processing and recurrent NNs are useful for time series processing. 
Although these models of deep neural networks are typical examples of deep learning \cite{LeCunNature}, the high performance of the models requires a large amount of computational cost. Meanwhile, reservoir computing has received considerable attention in terms of low computational cost, which is derived from recurrent NNs with a simple learning rule \cite{jaeger2001}. 

In reservoir computing, input signals are mapped into a high dimensional space represented by a large size of NN with randomly fixed weights. Moreover, the complex dynamics of NN preserves the memory of input signals in the high dimensional space so that it is possible to learn the desired output using a linear model. According to the property of fixed weights, the recurrent NN can be replaced other complex dynamic systems, namely physical reservoirs, such as a water surface filled in a bucket.

In this study, we consider a mathematical model, which exploits the dynamics of road traffic as a reservoir. 
The dynamics of road traffic are determined by the aggregation of individual vehicles while maintaining traffic order. Each vehicle is traveling a road network from its origin to its destination, and the network is basically common for all traveling vehicles in a limited space. Therefore, all the traveling vehicles generate the complex dynamics of traffic flow that can be used for reservoir computing. 

In the viewpoint of computational cost, a reservoir computing by measuring the real traffic flow and exploiting its complex dynamics can detect the computational resource not only from cyberspace, i.e., computers, but also from physical space. This means that the framework of reservoir computing with real traffic flow potentially reduces the computational cost of computers. 

For the purpose of this study, we verify the proposed concept of road traffic reservoir computing by simulating a mathematical model of traffic flow, before applying the idea to traffic dynamics in the real world. Precisely, we consider one of the simplest models of the traffic flow, in which some of the fundamental conditions for real traffic flow are assumed, and numerically simulate it to confirm the feasibility of the road traffic reservoir computing. Moreover, we simulate the multi-agent based model of traffic flow with the optimal velocity rule \cite{Bando1995} and verify the proposed concept in this model as well.

\section{Model}
First, to make the road network simple, we consider a $N \times N$ lattice network system with a traffic signal at the junctions. 
To introduce reservoir units into the system, each traffic signal has its internal state which is determined by the dynamics of traffic flow intersecting the junction of the traffic signal. 
Precisely, we define a phase $\theta_i$ of the traffic signal $i \in \{1,2,\ldots,N^2\}$ with the time constant $\tau_i$. Further, the dynamics of the phase $\theta_i$ is governed by the following discrete time dynamic systems.
\begin{eqnarray}
\theta_i(t+1)=\theta_i(t)+2\pi\frac{t}{\tau_i} +\xi_i,
\end{eqnarray}
 where $\xi_i$ is an initial phase shift. Now, we can define the internal state of the reservoir unit at the traffic signal $i$ (reservoir unit $i$) as follows:
\begin{eqnarray}
X^1_i(t)&=&u_i(t)\cos ^2 \theta_i(t),\\
X^2_i(t)&=&u_i(t)\sin ^2 \theta_i(t),
\end{eqnarray}
where the set of $\{X^1_i(t), X^2_i(t)\}$ is the internal state of the reservoir unit $i$ at time $t$, and $u_i(t)$ is a sum of inflow to the junction of the traffic signal $i$. The inflow to the junction is considered later. Note that the sinusoidal function for calculating internal states can be replaced with other nonlinear functions.

Next, we define the dynamics of traffic flow which satisfy the fundamental assumptions of the LWR (Lighthill, Whitham, Richards) model \cite{LWR}, i.e. 1) the conservation law for the number of vehicles and 2) the concave functional relation between flow $q$ (vehicles per hour) and density $k$ (vehicles per mile). 
For simplicity, we assume that the considered phase of flow and density is in that of free flow, namely there is no traffic jam in the model and $q$ is proportional to $k$. Due to the above assumptions, we define the dynamics of traffic flow as follows. The number of vehicles in the road between neighboring junctions $i$ and $j$ at time $t$ is denoted by $Q_{ij}(t)$. The length of the corresponding road is denoted by $L_{ij}$. Further, we define $u_{ij}(t)=Q_{ij}(t)/L_{ij}$ as the inflow to the node $i$ from the road between nodes $i$ and $j$. This means that the velocity is constant, or equal to 1, and the amount of inflow is governed by the density $k_{ij}(t)=Q_{ij}(t)/L_{ij}$. Therefore, $u_i(t)=\sum_{j\in A_i} u_{ij}(t)$, where $A_i$ is a set of neighboring nodes to $i$. We call this model a density model.

In addition, the outflow from node $i$ is determined by a sum of parts of inflows to the node, namely the amount of outflow with respect to one inflow is determined stochastically with the probability of $w_r$, $w_l$, and $w_s$, which means turning right, turning left, and going straight, respectively. Hence, $w_r+w_l+w_s=1$ is satisfied. 
These three probabilities are assigned to all inflows for one node randomly. 
In the case of a boundary in the lattice network, we do not consider the output and input of a vehicle for simplicity. 
Finally, we introduce a stop and go function to the traffic signal $i$ such that the inflow from one direction is zero (i.e. {\it stop}) and its perpendicular direction has some value (i.e. {\it go}) when $0\le\mod(\theta_i,2\pi)<\pi$ at the junction $i$. Otherwise, the former direction is ``go'' and the latter is ``stop''. 
Note that the dynamics of traffic flow in this model are determined by the inflow, the outflow, and the phase.

Regarding the readout from the reservoir units, we use a standard linear model. In other words, the output weight of the model $y=W^\text{out}\mathbf{X}$ is estimated by the ridge regression as $W^\text{out}=\mathbf{Y}\mathbf{X}(\mathbf{X}\mathbf{X}^\top+\beta \mathbf{I})^{-1}$. 
$\mathbf{Y}$ and $\mathbf{X}$ are the $L$-length time series of a teacher signal $y(t)$ and the states of the reservoir units $\mathbf{X}(t)$, respectively. We define $\mathbf{X}(t)=[1;U(t);\mathbf{X}^1(t);\mathbf{X}^2(t)]$ (def. (a)) or $\mathbf{X}(t)=[1;\mathbf{X}^1(t);\mathbf{X}^2(t)]$ (def. (b)). Here, $\mathbf{X}^1(t)=[X^1_1(t), X^1_2(t),\ldots,X^1_N(t)]$ and the same as $\mathbf{X}^2(t)$. Further, the dimension of $\mathbf{X}$ is $(1+1+2N) \times L$ for def. (a) and $(1+2N)$ for def. (b). $\beta$ is a hyper parameter. 
Note that $U(t)=y(t-T)\in \mathbb{R}$ corresponds to the input signal for the reservoir however, is not necessarily added to the reservoir units depending on a task. We discuss this point later. $T$ is forecast horizon explained in the next section. $[\cdot;\cdot]$ is a vector concatenation. 

In addition, we also consider the multi-agent based model of traffic flow in which each agent follows the vehicle ahead on the same road with the optimal velocity model \cite{Bando1995}. Depending on the parameters, a traffic jam can be observed in this model. Therefore, if the prediction on this model is performed well, it should be more realistic than the density model. We call this model a multi-agent model. In this model, the above def. (a) is slightly modified as: $\mathbf{X}(t)=[1;U(t);\mathbf{X}^1(t);\mathbf{X}^2(t);\mathbf{K}(t-T)]$, where $\mathbf{K}(t-T)\in \mathbb{R}^M$ is a vector of the densities in $M$ roads at time $t$.
\vspace{-1mm}
\section{Prediction task}
In this study, we consider the following two prediction tasks. \vspace{-1mm}
\begin{enumerate}
\setlength\itemsep{1mm}
\item Prediction of the density in a link of the network with $T$ step forecast horizon.
\item Prediction of temperature in Tsukuba city with $T$ step forecast horizon.  
\end{enumerate}\vspace{-1mm}
The first task is predicting an internal state of the traffic dynamics so that there is no external input to the system. Moreover, the second task is based on a prediction of an arbitrary time series independent from the traffic dynamics. Therefore, an external input needs to be added to the reservoir units. We consider how to apply an external input $u(t)$ to the system as follows. Note that $u(t)=y(t-T)$ in this task. At every time step, the phase $\theta_i(t)$ is perturbed by the input as $\theta_i(t+1)=\theta_i(t)+2\pi\frac{t}{\tau_i} +\xi_i+W^\text{in}_i u(t)$, where $W^\text{in}$ is a random vector. Regarding the definition of $\mathbf{X}$, the defs. (a) and (b) are applied to the tasks (1) and (2), respectively. Note that the def. (b) makes the prediction easier for task (2), so we do not consider this case. 

In addition, we consider a constraint condition that all the reservoir units are not necessarily used for the prediction task 1. Here, we introduce a parameter $p\in[0,1]$ representing the fraction of available reservoir units out of all units. This condition is introduced to evaluate the ability of predicting traffic density in a target road that is not connected to reservoir units used for the prediction. We investigate how many reservoir units can be reduced for good prediction performance. 

In the following section, we numerically evaluate the precision of the prediction tasks 1 and 2 by the reservoir units with the density model. Regarding the multi-agent model, we check the prediction ability of the reservoir system for task 1. 
The precision is evaluated by the logarithm of the normalized root mean-square error (logNRMSE):
$\displaystyle
E(y,\hat{y}) = \log(\sqrt{\langle \|y-\hat{y} \|^2 \rangle / \langle \|y - \langle y \rangle \|^2\rangle)}, 
$
where $\langle \cdot \rangle$ represents the time average, and $\hat{y}$ is the predicted $y$.

\section{Results}
\begin{figure}[ht]
\begin{minipage}{0.45\hsize}
\centering
\includegraphics[scale=0.2]{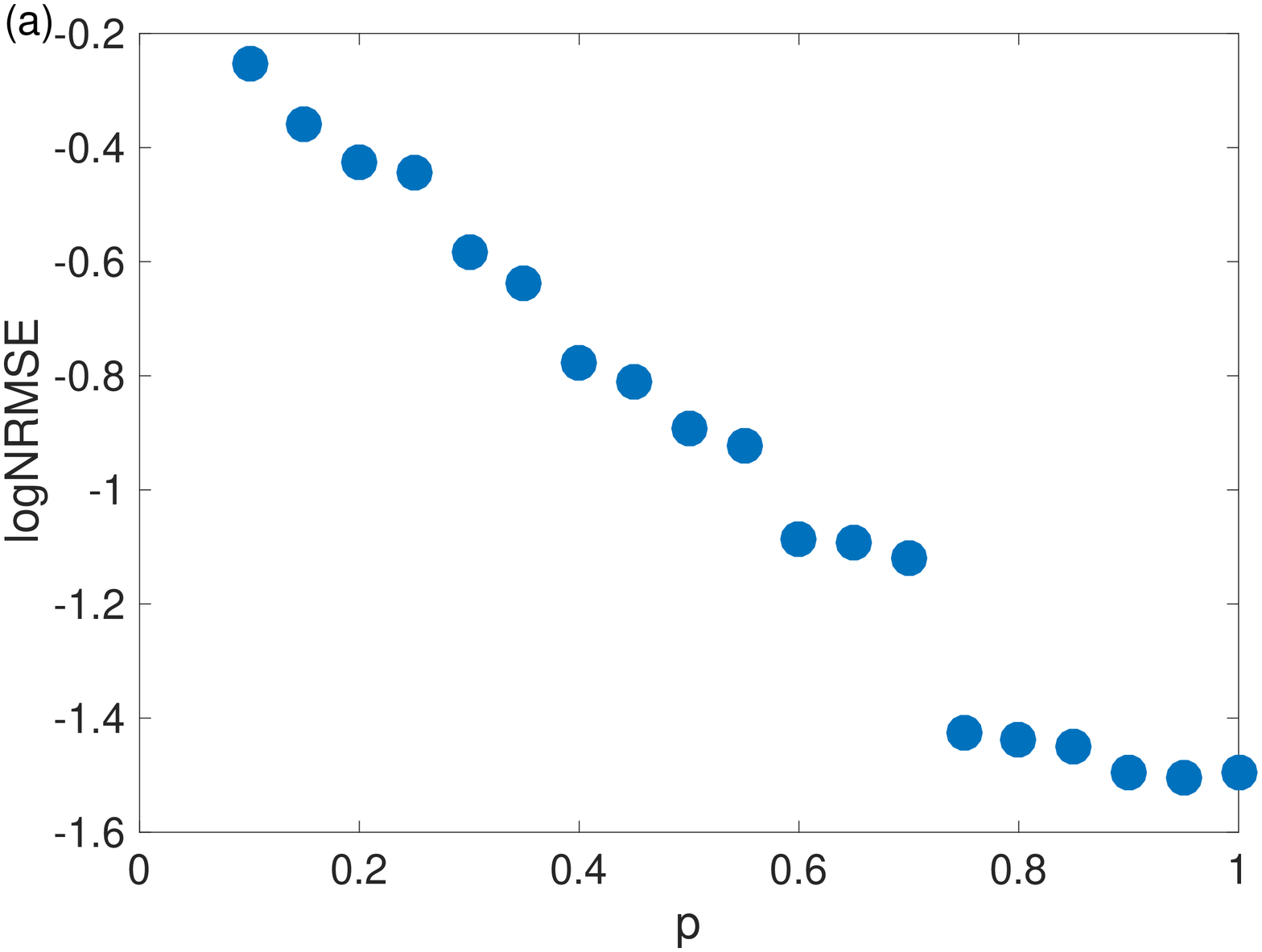}
\end{minipage}
\begin{minipage}{0.45\hsize}
\centering
\includegraphics[scale=0.2]{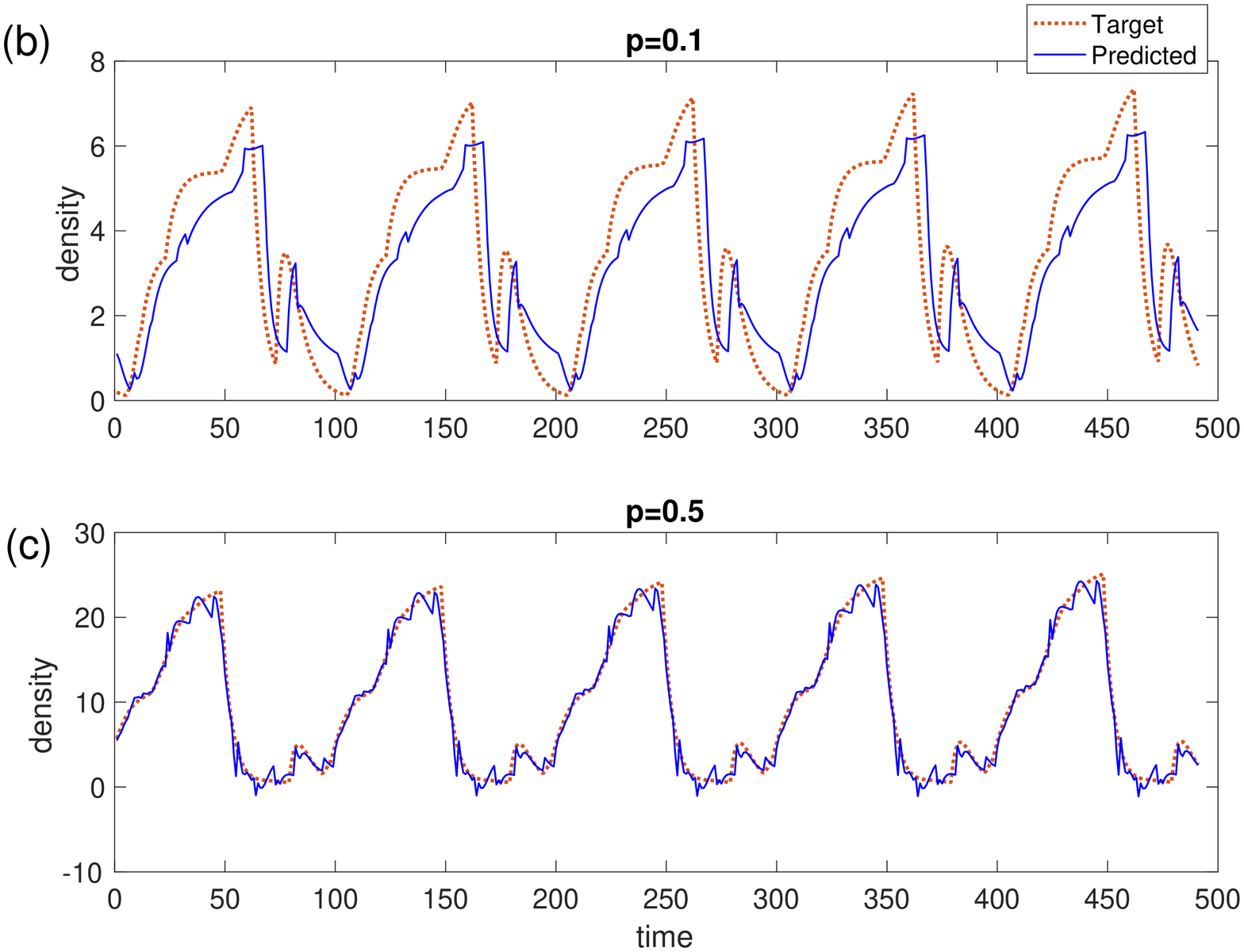}
\end{minipage}
\caption{(a) Precision of prediction in the density model with respect to the fraction parameter $p$;(b) The size of the road network $N=5$;(c) Predicted time series of the density in one road that is not connected to the reservoir units for prediction. }\label{Fig:dm}
\end{figure}

In this section, we show the numerical results of prediction for tasks 1 and 2. 
First, Fig. \ref{Fig:dm} (a) shows the precision of prediction depending on the fraction parameter $p$ for task 1 when $N=5$, $T=5$, $\forall i,\tau_i=100$, and $\beta=10^{-8}$. The results are averaged over 20 trials. The lengths of training and test time series are $4000$ and $2000$ time steps, respectively.
It is clearly observed that precision increases with an increasing value of $p$. 
This dependency is reasonable, as reducing the number of reservoir units makes the prediction worse. In fact, delayed prediction for very small $p$ is observed, as shown in Fig. \ref{Fig:dm} (b). It is not indicated that this is a prediction, as a predicted signal generates $T$ step ahead of the teacher signal with some scale transformation. Otherwise, it is possible to predict traffic density sufficiently well for an intermediate value of $p$, say $p=0.5$, as shown in Fig. \ref{Fig:dm} (c). This implies that road traffic in some areas can be exploited as reservoir units for predicting the traffic density in other areas. Note that the delayed prediction is also observed even for a large value of $p$, when $\tau_i$ are varied and the time series of the density is complex. 

\begin{figure}[ht]
\begin{minipage}{0.45\hsize}
\centering
\includegraphics[scale=0.2]{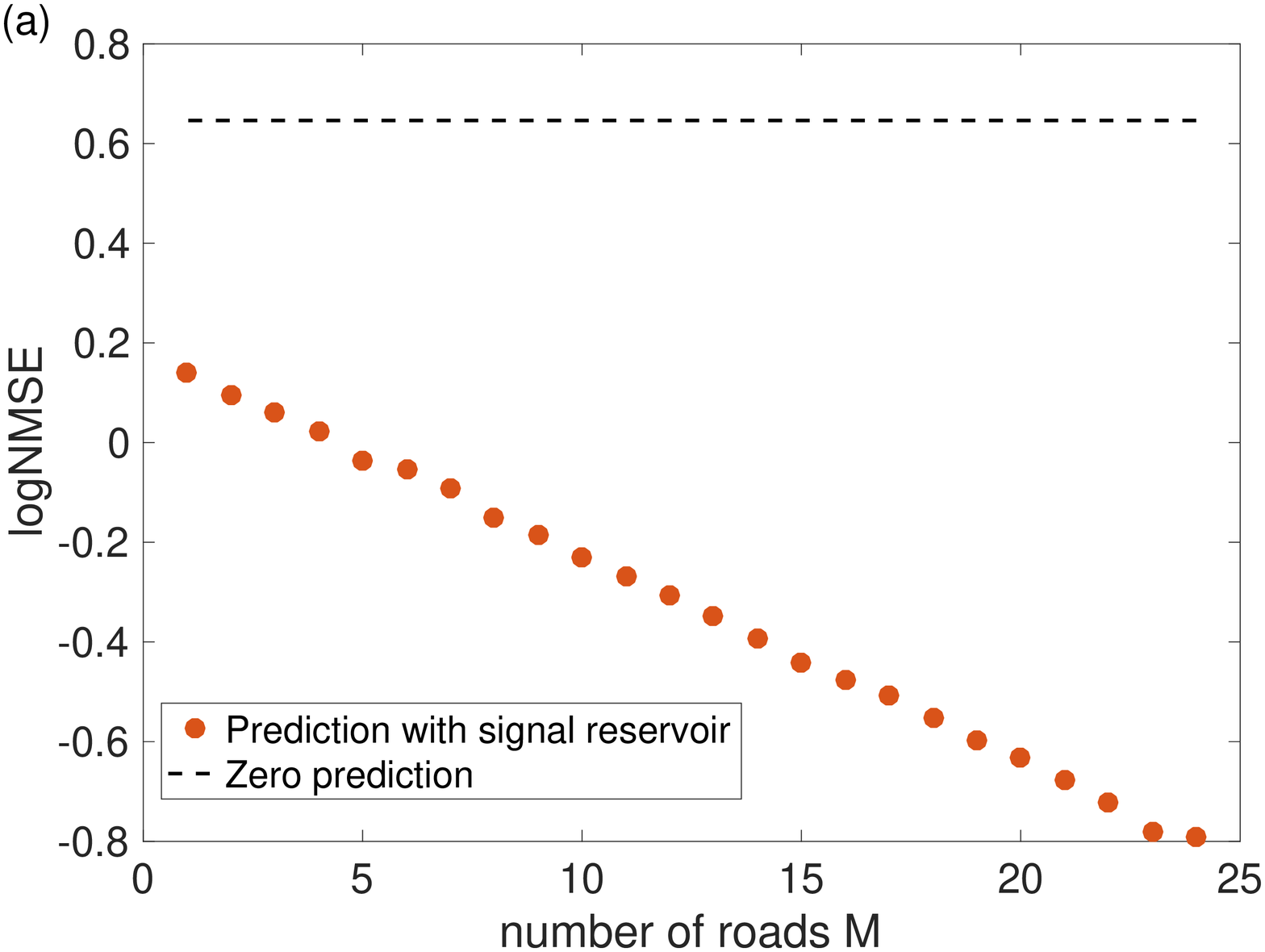}
\end{minipage}
\begin{minipage}{0.45\hsize}
\centering
\includegraphics[scale=0.2]{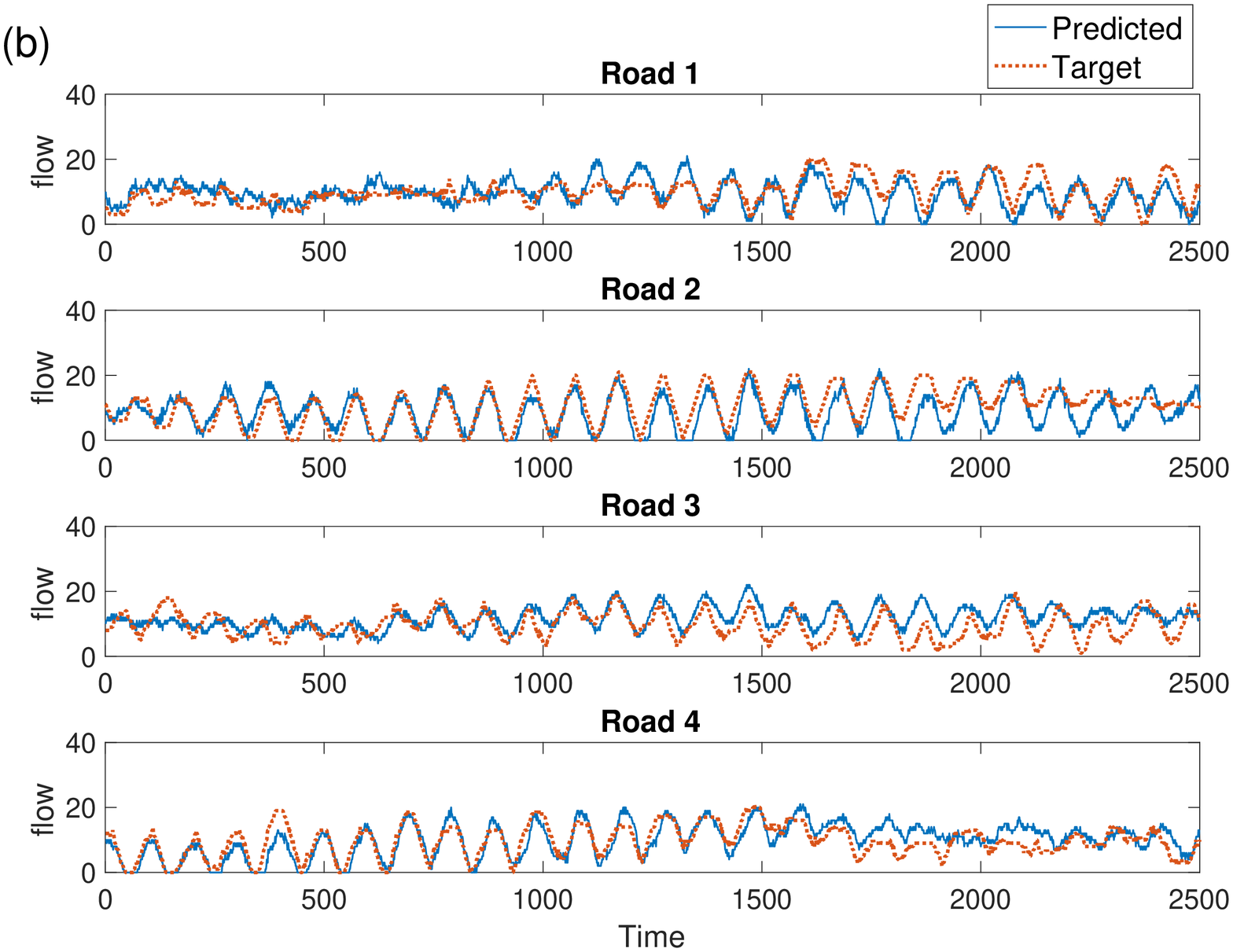}
\end{minipage}
\caption{(a) Precision of prediction in the multi-agent model with respect to the number of roads $M$. $M$ roads are determined randomly from all $24$ roads for $N=3$. The results are averaged over 10 trials; (b) The predicted time series of densities for 4 omitted roads from the estimation of $W^\text{out}$ when $M=20$. }\label{Fig:mab}
\end{figure}

In Fig. \ref{Fig:mab}, we show the results of task 1 for the multi-agent based model with $N=3$. Fig. \ref{Fig:mab} (a) shows the dependency of precision on the number of roads $M$ that determine the dimensions of $\mathbf{K}(t)$. The results are averaged over 10 trials. The lengths of training and test time series are $2500$ and $2500$ time steps, respectively. As shown in the figure, precision decreases with decreasing $M$. However, it is not necessarily required for the density of all roads, i.e., $M\le24=4N(N-1)$ is sufficient for prediction in some cases. In fact, Fig. \ref{Fig:mab} (b) shows the time series of prediction in the case of $M=20$, where 4 roads are omitted from estimating $W^\text{out}$. As shown in the figure, the densities of the 4 roads are predicted moderately from the densities of the other 20 roads and the 9 reservoir units. 
These results imply that our concept of road traffic reservoir computing is applicable in the case that the traffic flow in a road network includes both jam flow together with free flow. 

\begin{figure}[ht]
\begin{minipage}{0.45\hsize}
\centering
\includegraphics[scale=0.2]{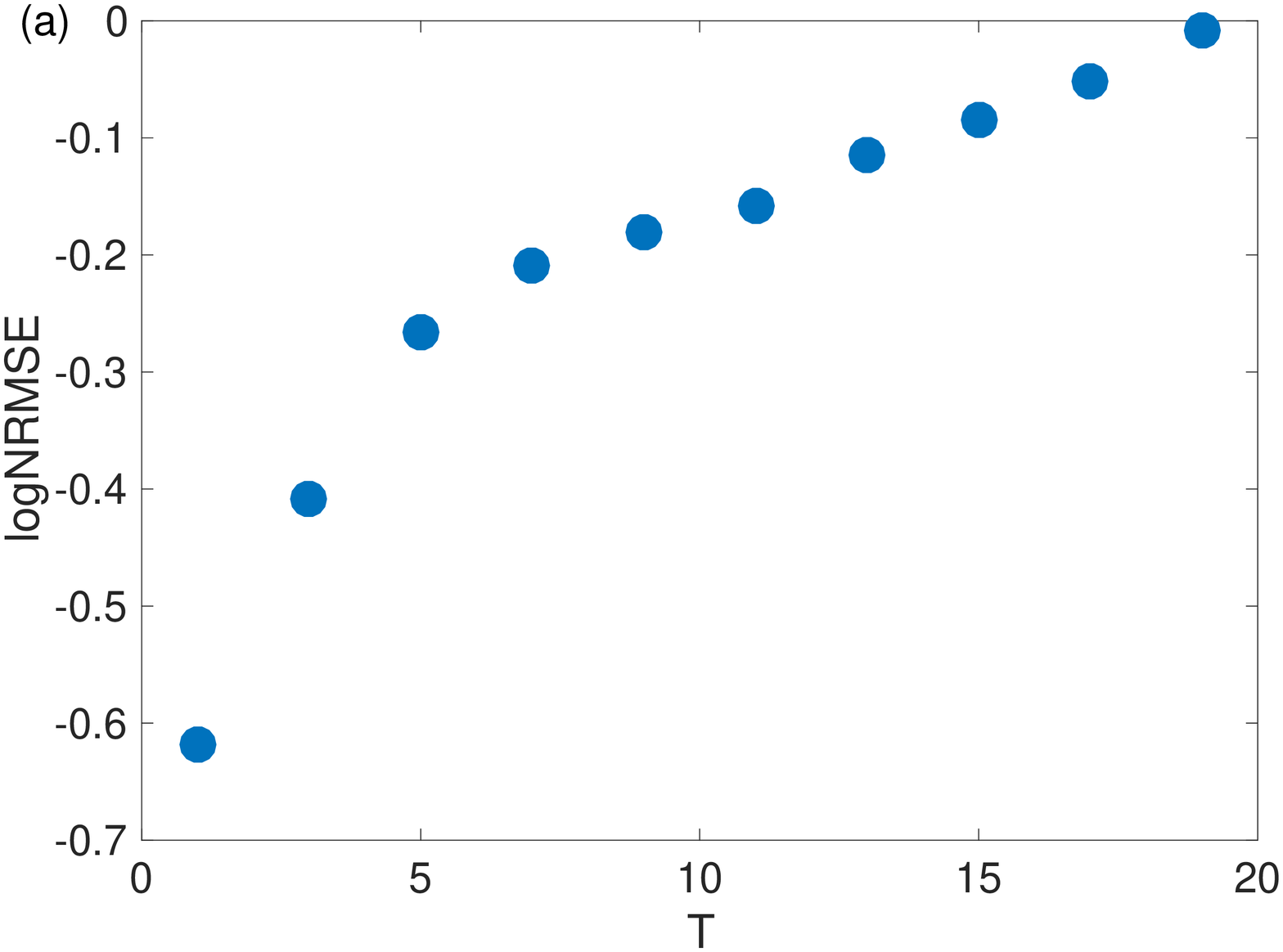}
\end{minipage}
\begin{minipage}{0.45\hsize}
\centering
\includegraphics[scale=0.2]{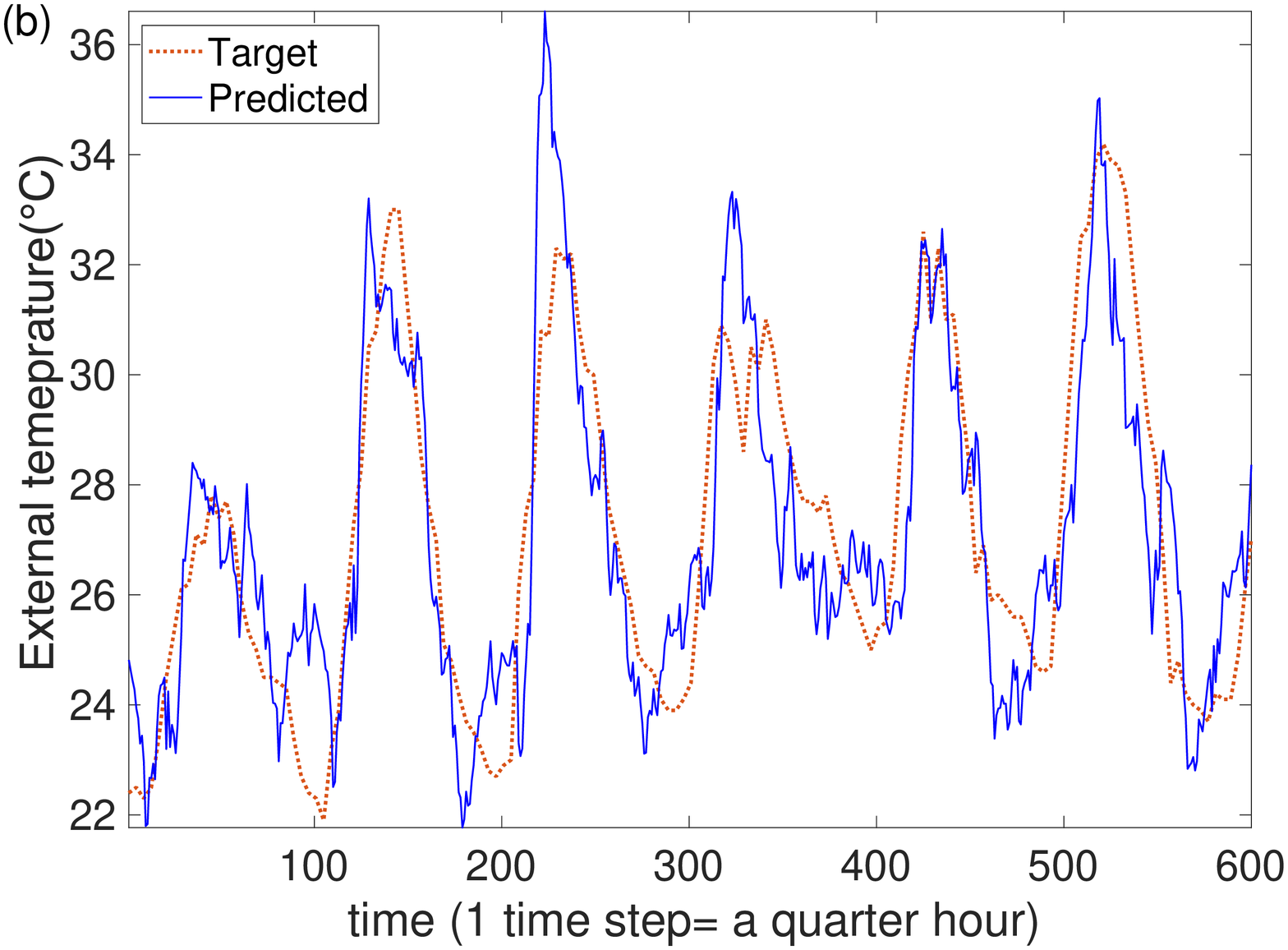}
\end{minipage}
\caption{(a) Precision of predicting temperature in Tsukuba city by the density model with respect to the forecast horizon $T$. The size of road network $N=10$. The results are averaged over 20 trials. (b) Predicted temperature time series with $T=10$. The time series are the average of five trials.}\label{Fig:tsukuba}
\end{figure}

Fig. \ref{Fig:tsukuba} shows the prediction results of task 2 by the density model. 
Fig. \ref{Fig:tsukuba} (a) shows the dependency of precision on the forecast horizon $T$. The lengths of training and test time series are $5000$ and $600$ time steps, respectively. Precision decreases with increasing $T$. For large $T$, the delayed prediction is also observed in this task. 
As one-time step corresponds to 15 min in this case, it is possible to take a sufficiently large forecast horizon with moderate prediction. In fact, Fig. \ref{Fig:tsukuba} (b) shows an example of the predicted time series for $T=10$ (150 mins.), which is the average of five trials. This averaging operation is reasonable for real time prediction because the estimation of $W^\text{out}$ can be done in parallel for $W^\text{in}$.

\section{Conclusions and Future works}
In this study, we proposed a concept of reservoir computing with road traffic dynamics and verified it by numerical simulations for the flow-density model as well as the multi-agent model with the optimal velocity rule. We summarized the characteristics of a road traffic reservoir by computing as follows. i) In the task of predicting the internal state of the system, it does not require external inputs to the system, which implies that the system is closed. ii) It is possible to observe the learning process in the internal state. Therefore, we may discuss the interpretability of prediction deductively. iii) Once real time traffic data are available, a part of the prediction calculation is assigned to the physical phenomena, which implies that the computational cost can be reduced. 
As for future research, we will systematically simulate the proposed models to make the relation between the multi-agent model and the density model clear. Likewise, the problem of delayed prediction that is observed for a complex time series can be solved by improving the linear model for output from the reservoir units.


\begin{footnotesize}





\end{footnotesize}


\end{document}